\documentclass[conference]{IEEEtran}

\usepackage{tikz}
\usetikzlibrary{decorations.pathreplacing,calc}
\usepackage{amsmath,bm,bbm,amsthm, amssymb, authblk}
\usepackage{caption}
\usepackage{subcaption}
\newcommand{\E}{\mathbb{E}} 

\newcommand{\R}{\mathbb{R}}

\def\eq{\begin{equation}}
\def\en{\end{equation}}
\def\laI{\lambda_{\rm I}}

\def\laG{\lambda_{\rm G}}

\newtheorem{theorem}{Theorem}[section]

\newtheorem{lemma}[theorem]{Lemma}

    \def\a{\alpha}

    \def\phi{\varphi}
    \def\g{\gamma}
    \def\la{\lambda}

    \def\x{\xi}

    \def\L{\Lambda}

    \def\P{{\Phi}}
    
    \def\T{\T}

    \def\V|{{\Vert}}

    \def\d{{\rm d}}
    \def\E{\mathbb{E}}
    \def\V{\mathbb{V}}

    \def\ms{\mathsf}

    \def\P{\mathbb{P}}
    \def\R{\mathbb{R}}

    \def\vG{g}

\title{Malware propagation in urban D2D networks}

\author[1]{A.~Hinsen~\thanks{hinsen@wias-berlin.de}}
\author[1]{B.~Jahnel~\thanks{jahnel@wias-berlin.de}}
\author[2]{E.~Cali~\thanks{elie.cali@orange.com}}
\author[2]{J.-P.~Wary~\thanks{jeanphilippe.wary@orange.com}}

\affil[1]{Weierstrass Institute for Applied Analysis and Stochastics, Mohrenstra{\ss}e 39, 10117 Berlin, Germany}
\affil[2]{Orange Labs, 44 Avenue de la R\'epublique, 92320 Ch\^atillon, France}

\begin{document}

\maketitle

\begin{abstract}
We introduce and analyze models for the propagation of malware in pure D2D networks given via stationary Cox--Gilbert graphs. Here, the devices form a Poisson point process with random intensity measure $\lambda\Lambda$, where $\Lambda$ is stationary and given, for example, by the edge-length measure of a realization of a Poisson--Voronoi tessellation that represents an urban street system. We assume that, at initial time, a typical device at the center of the network carries a malware and starts to infect neighboring devices after random waiting times. Here we focus on Markovian models, where the waiting times are exponential random variables, and non-Markovian models, where the waiting times feature strictly positive minimal and finite maximal waiting times. We present numerical results for the speed of propagation depending on the system parameters. 

In a second step, we introduce and analyze a counter measure for the malware propagation given by special devices called white knights, which have the ability, once attacked, to eliminate the malware from infected devices and turn them into white knights. Based on simulations, we isolate parameter regimes in which the malware survives or is eliminated, both in the Markovian and non-Markovian setting. 
\end{abstract}


\section{Introduction}
%
%
Present day telecommunication networks are ill equipped for the rapidly growing demand for mobile data transfers. 
With the fifth generation of mobile networks, paradigmatic shifts in the design of the network are on the agenda.
A critical aspect here is the r\^ole of infrastructure. Multilayered cellular networks with possible incorporation of relaying mechanisms are under investigation not only in the scientific community~\cite{baccelli2009stochastic1, baccelli2009stochastic2, Wu,JaKo20}, but also in industry~\cite{3GPPRelay}. All these new designs have in common  a rapid increase of degrees of freedom in the system. The central r\^ole of base stations is reduced in favor of an increasingly important r\^ole of relays. In particular, also the users of the system will be attached a relay functionality in the system. As a result, the network becomes more and more decentralized. 
Exploring the possible benefits of such new architectures is in full swing in the academic and the industrial research. For a survey on device-to-device (D2D) communication in cellular networks see for example~\cite{asadi2014survey}.

%
%
Of crucial importance in all these future scenarios with less centralized architectures is a good understanding of vulnerability and security, in particular of the way in which malware (e.g., proximity-based propagation sabotage software or computer killing viruses like Cabir or CommWarrior) spreads in such networks \cite{Valler2011}. For usual networks, a number of strategies have been exploited by operators in order to restrict the spread of the malware and to keep the functionality of the network available \cite{Xie}. For security in D2D communication networks see the review \cite{Wang}. However, the new challenges accompanying the system decentralization also include the question how successful these defense strategies can be in such systems, in particular since the spread of malware (more generally of any information) follows a different set of rules than in centralized networks. 

In this paper, we introduce a framework for the modeling of malware spreading in an urban D2D scenario. One of the main parts of this framework is that the high complexity of the real-world system is reflected by a probabilistic perspective. Configurations of devices in the cities appear as random configurations of points or vertices, and their connectivity structure is represented by edges between these vertices. Similarly, the times at which data is transmitted along edges to neighboring devices are also assumed random, which results in a stochastic process of interacting devices. In order to find a balance between describing these complexities and being able to interpret the theoretical and numerical findings, we will mainly focus on models that can be described by a set of parameters of reasonable size.

We analyze numerically the speed of the infection spread based on the system parameters, for which we introduce two different updating mechanisms. We
also evaluate the possibility of tackling the malware by the introduction of {\em white knights} into the system. The success of this counter measure is then measured via its effect on the survival of the malware over long times and large areas. 
Competition models of this type have been considered in the literature for various types of fixed network models and in a Markovian setting \cite{Ko05,Bo14,Ko15,Ko16,DuJuTa18,HiJaCaWa19}. However, the present paper is devoted to the study of such models on particularly realistic and random network models, also going substantially beyond the Markovian setting for the propagation model.

\section{Model}
%
%
\subsection{Street-system based network models}
\begin{figure}[!htpb]
\centering
\input{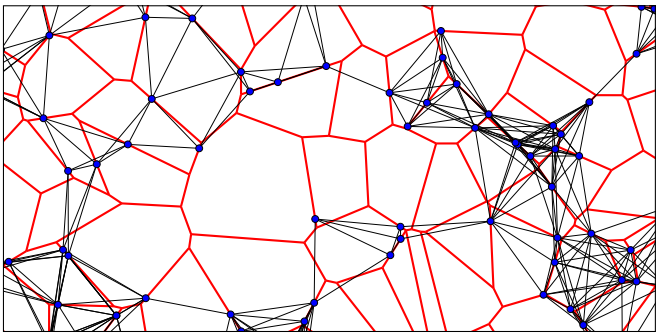}
\caption{Realization of randomly placed devices (blue) confined to a street system given by a Poisson--Voronoi tessellation (red). Edges (black) are drawn whenever two devices are at sufficiently close proximity of each other.}
\label{Street}
\end{figure}

For the distribution of devices in space we adopt the stochastic geometry model of \emph{Cox point processes} and later base our simulations on particular choices of environment measure. More precisely, a Cox point process in $\R^2$ is a Poisson Point process with random intensity measure $\la\Lambda$, where the directing measure $\Lambda$ represents a stationary ergodic random environment. For example, starting from a homogeneous Poisson--Voronoi tessellation $S$, we define the (random) measure $\L(A)=|S \cap A|$ which assigns to any measurable subset $A\subset\R^2$ the street length in $A$, see red cells in Figure~\ref{Street}. 
This modeling ansatz is supported by some statistical analysis performed in~\cite{courtat2012promenade}, at least for European cities. (Non-European cities with more rectangular street systems have been analyzed for example in~\cite{HiHiJaCa18}.) In this example, since the underlying homogeneous Poisson point process that is used to create the Poisson--Voronoi tessellations can be parametrized by one parameter, the density of streets can also be described by one parameter, $\g = \E|S \cap [-1/2,1/2]^2|$, the {\em expected street length in a unit area}. In our applications, the quantities of interest will usually only depend on the parameter $\g$, rather than on the whole distribution of the street system.

To finish the description of a Cox point process, for a given street system $S$, we assume that the devices form a Poisson point process $X = \{X_{i}\}_{i\ge1}$ with intensity measure $\la|S\cap \d x|$, see blue dots in Figure~\ref{Street}. Here, $\la > 0$ is another system parameter describing the \emph{linear intensity of devices} on the streets.

For the network model, we use the classical \emph{Gilbert graph $\vG_{r}(X)$ with connection radius $r>0$}. That is, devices $X_{i}, X_{j}\in X$ are connected by an edge in $\vG_{r}(X)$ if and only if $|X_{i} - X_{j}|\le r$, see thin black lines in Figure~\ref{Street}. 

\medskip
The network model thus depends on the three central parameters $\la$, $\g$ and $r$, which correspond to fundamental characteristics of real-world wireless networks and obey certain scaling relations. For example, the product $\la\gamma$ describes the spatial intensity of devices, i.e., the expected number of devices per unit area. 
Not all parameters are equally accessible from an operator's perspective. For example, the connection radius $r$ depends very much on technology (e.g., LTE, Bluetooth, or WIFI). On the other hand, the street intensity $\gamma$ is given, but it is different from one city to the other. Hence, results should cover a certain range of values. Similarly, 
the linear intensity $\la$ depends on the number of devices participating in the system and thus might vary substantially. 

Next, we introduce malware to the system and describe how its propagation can be formulated. 

\subsection{Malware propagation models}
%
%
So far, devices $X_i$ in the system carry no information about being infected or not, and there is no time component involved. We start by labelling each device $X_i\in X$ and write $\xi(t,X_i)$ for the state of the device $X_i$ at time $t\ge 0$. In the first step, we only want to distinguish two possible states, namely $\rm S$ and $\rm I$, for \emph{susceptible} and \emph{infected}, i.e., $\xi(t,X_i) \in \{\rm S, \rm I\}$:
\begin{align*}
\xi(t,X_i)=
\begin{cases} {\rm S} &\hspace{-.4cm}\text{ : device } X_i \text{ is susceptible at time }t \\ 
{\rm I} &\hspace{-.4cm}\text{ : device } X_i \text{ carries malware at time } t.
\end{cases}
\end{align*} 
We then define the set of infected devices at time $t$ by 
\begin{align*}
I(t)=\{X_i\in X\colon \xi(t,X_i)={\rm I}\}.
\end{align*}
We are interested in the collective behavior of a large number of interacting devices given by the vertices of $\vG_{r}(X)$ as presented above. 
Our general modeling assumption is that a change of status of $X_i\in X$ happens at random times due to fluctuations in the system, and these times depend on the configuration in the vicinity of $X_i$. Systems of this type are often called \emph{interacting particle system} and a large body of literature is available typically for non-random particle positions and exponential waiting times, see for example~\cite{Li85,Li13,KeDu12,Sp71}.

In view of the application to malware propagation in realistic D2D networks, in this paper, we use an extended framework where instead of fixed geometries, the underlying geometric model is random and given by $\vG_{r}(X)$. Further, besides the usual exponential waiting times that we introduce next, we also study models with non-exponential waiting times, at least via simulations, see Section~\ref{results}. 

In order to introduce interaction into our system, we define the jump rate of a device $X_i$ to jump from a state $\mathrm S$ to a state $\mathrm I$ as $\lambda_{\rm I}>0$ times the number of infected neighbors of $X_i$ in $\vG_{r}(X)$. At this point there are no defense mechanisms at work and thus we assume that infected devices cannot become susceptible again. Using the power expansion of the exponential and the Landau notation, we define
\begin{align*}
&\P(\xi(t+h,X_i)={\rm I}\mid\xi(t,X_i)={\rm S})=\\
&h \lambda_{\rm I}\#\{X_j\in X:\, |X_j-X_i|<r\text{ and } \xi(t,X_j)={\rm I}\} + o(h).
\end{align*}
Here, the scalar parameter $\lambda_{\rm I}$, the {\em infection rate}, can be used to adjust the speed of the microscopic updates. 
We call this model the {\em Markovian $\rm SI$-model}. In the literature it is also referred to as the {\em Richardson model}, see~\cite{Ri73}. 
We have the following well definedness result.
\begin{lemma}\label{Prop1}
For almost-all Gilbert graphs $g_r(X)$ we have that $\#I(0)<\infty$ implies $\#I(t)<\infty$ for all $t\ge 0$. 
\end{lemma}
\begin{proof}
Let $\tau_n=\inf\{s\ge 0\colon I(s)\not\subset B_n\}$, where $B_n=[-nr,nr]^2$, then it suffices to show that $\tau_n\uparrow\infty$ almost surely. The proof rests on coupling arguments for the propagation process and ergodicity of the random environment. First, note that 
\begin{align*}
\tau_n=\sum_{i=1}^n\tau_{i}-\tau_{i-1}\ge \sum_{i=1}^n\tau'_i,
\end{align*}
where $\tau'_i\sim \ms{Exp}(\laI\sum_{X_j\in X\cap B_i\setminus B_{i-1} }\deg (X_j))$ represents the fastest possible crossing of the annulus $B_i\setminus B_{i-1}$. Here $\deg (X_j)$ denotes the degree of $X_j$ in $g_r(X)$. 

Next, due to ergodicity, there exist $a,b>0$ and $m=m(g_r(X))$ such that for all $n>m$ we have 
\begin{align*}
\sum_{X_j\in X\cap B_i\setminus B_{i-1}} \deg (X_j)<a |B_i\setminus B_{i-1}|\le 4ari
\end{align*}
for at least $bn$-many annuli $i$. There, $ \E[\tau'_i] >\frac{1}{\lambda 4ari}$ holds. This implies that 
\begin{align*}
\lim_{n\uparrow\infty}\E[\tau_n]\ge\lim_{n\uparrow\infty} \sum_{i=0}^n\E[\tau'_i] = \infty,
\end{align*}
Using similar arguments we can show for the variance that $\lim_{n\uparrow\infty}\V[\sum_{i=0}^n\tau'_i]<\infty$ and hence $\tau_n\uparrow\infty$ almost surely. 
\end{proof}

Although the Markov property has substantial advantages for the analytical approach to malware propagation, the implied description via exponential waiting times is a restriction and also does not reflect some of the details of a realistic transmission process. Since, from a simulation perspective, the Markovian approach is not a substantial advantage, in this section, we develop an extended framework for interacting particle systems based on more general waiting-time distributions. The distributions that we have in mind feature a deterministic initial time step where transmission is impossible and a maximal waiting time until which transmissions must be finished. Further, here statistical input can be incorporated, which might propose certain bell-like shapes for the density of the waiting times.
In the simplest case, we can work with a distribution where the infection attempt happens at times which are uniformly distributed in $[c_1,c_2]$ with $0<c_1<c_2$, i.e., we have a density $f(t)=(c_2-c_1)^{-1}\chi_{[c_1,c_2]}(t)$ for the iid renewal times. 

We call this model the {\em non-Markovian {\rm SI}-model}. We have the following well-definedness result. 
\begin{lemma}\label{Prop2}
For the propagation model with iid renewal times that have a density with support of the form $[c_1,c_2]$ for some $0<c_1<c_2$, we have that for almost-all Gilbert graphs $g_r(X)$, $\#I(0)<\infty$ implies $\#I(t)<\infty$ for all $t\ge 0$. 
\end{lemma}
\begin{proof}
Since in any finite time window, there can only be at most $t/c_1$ consecutive jumps and hence the infection can reach at most distance $rt/c_1$. 
\end{proof}

\begin{figure}[!htpb]
\includegraphics[width=1\linewidth]{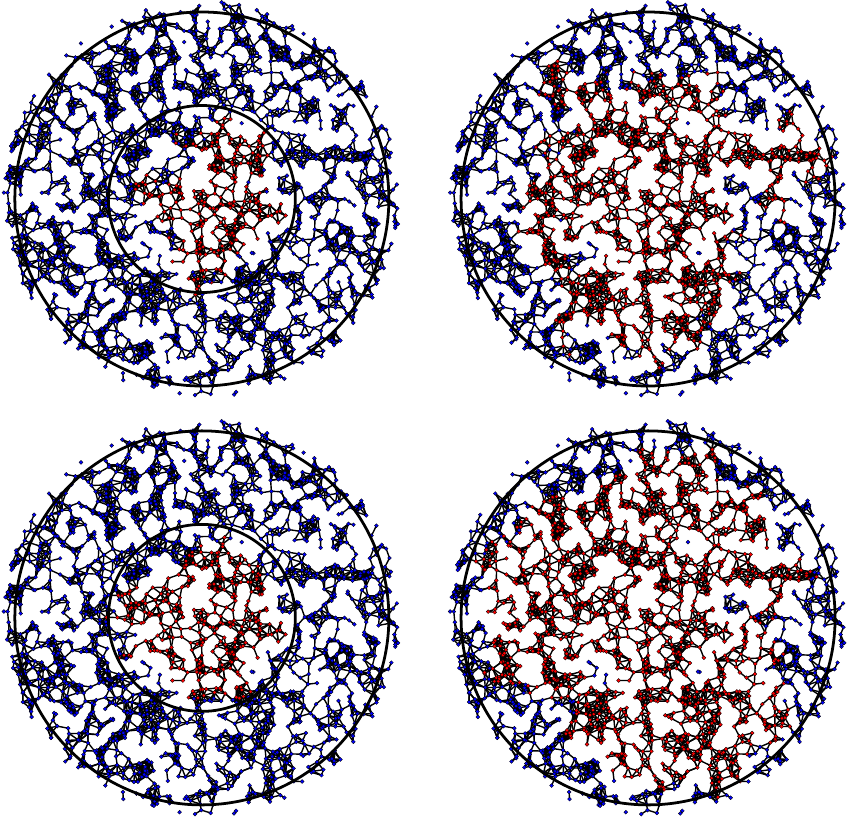} 
\caption{Realization of randomly placed devices on a street system of Poisson--Voronoi tessellation type with $\gamma=20\,\rm km^{-1}$, $\la=1.2\, \rm devices/km$ and $r=300\,\rm m$. Upper row: The Markovian {\rm SI}-model with parameter $\lambda_{\rm I}=1$ stopped at the time at which the malware has reached the radius $u=2.5\,\rm km$ (left) and $u=5\,\rm km$ (right), indicated in black. Lower row: The non-Markovian {\rm SI}-model with uniform waiting times on $[40\, \rm sec, 120\, \rm sec]$ stopped at the time at which the malware has reached the radius $u=2.5\,\rm km$ (left) and $u=5\,\rm km$ (right), indicated in black.}
\label{Pix-Rich-Stop-Exp}
\end{figure}

In Figure~\ref{Pix-Rich-Stop-Exp} we present pairs of snapshots for the {\rm SI}-model with exponential waiting times and uniform waiting times on $[c,d]$, all in a finite ball. The simulations are stopped at the stopping times at which a certain radius is reached. 

\subsection{White-knights counter measure}
%
%
Let us introduce a counter measure based on the presence of another type of software, which permanently eliminates malware. For this we introduce a new state $\rm G$ and thus increase the local state space to $\{{\rm S, I, G}\}$. Updates can only be performed in this order 
$${\rm S} \rightarrow {\rm I} \rightarrow {\rm G}.$$ 
In words, susceptible devices, once infected, cannot become susceptible again, and only infected devices can become patched. The reason why the software patch cannot be transmitted to susceptible devices at once is the following. Due to privacy regulations, only after an infection attempt, the infection becomes known to other devices and only then retaliation is authorized without explicit consent of the infected device. The software patch corresponds to an immunization of the device and thus $\rm G$ is an absorbing state. We call devices in the state $\rm G$ {\em white knights}. 

In order to have a Markovian structure, we assume exponential waiting times for the patch installation with parameter $\laG>0$, the {\em patch rate}. This leads to the following description via generators
\begin{align*}
&\P(\xi(t+h,X_i)={\rm I}\mid\xi(t,X_i)={\rm S})= \\
&h \laI\#\{X_j\in X:\, |X_j-X_i|<r\text{ and } \xi(t,X_j)={\rm I}\} + o(h),\\
&\P(\xi(t+h,X_i)={\rm G}\mid\xi(t,X_i)={\rm I})= \\
&h \laG\#\{X_j\in X:\, |X_j-X_i|<r\text{ and } \xi(t,X_j)={\rm G}\}  + o(h).
\end{align*}
We call this process the {\em Markovian {\rm SIG}-model}. 

Typically, in our analysis, we will adopt a setting where initially the infection is present at one typical node in the network, which we assume to be the origin. This reflects the idea that the malware is brought into the system by one device and we start our clock at that time. On the other hand, white knights have to be present in the system already at time zero. Their locations are random, more precisely, white knights are assumed to have the same spatial distribution as any other device. In other words, at initial time, we will assume that white knights form a Poisson point process $Y = \{Y_{i}\}_{i\ge1}$ on the street system $S$, independent of the other devices $X = \{X_{i}\}_{i\ge1}$. Its intensity measure is then given by $\rho|S\cap \d u|$, where $\rho > 0$ is another system parameter describing the \emph{linear intensity of white knights} on the streets. We note that alternatively, a Cox process with joint intensity $(\rho+\la)|S\cap \d u|$ can be realized and in a second step, devices can be labelled as white knights with iid probability $\rho/(\rho+\la)$.

\begin{figure}[!htpb]
\includegraphics[width=1\linewidth]{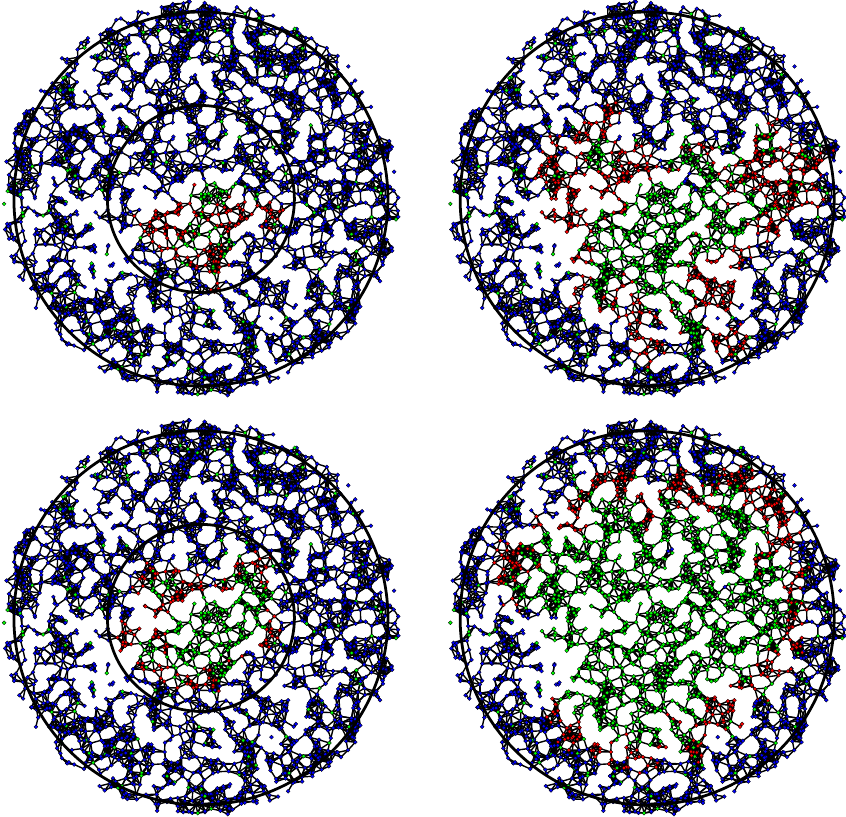} 
\caption{Realization of randomly placed devices on a street system of Poisson--Voronoi tessellation type with $\gamma=20\,\rm km^{-1}$, $\la=1.3\, \rm devices/km$, $\rho=0.1\, \rm devices/km$ and $r=300\,\rm m$. Upper row: The Markovian {\rm SIG}-model with parameter $\laI=1$ stopped at the time the malware has reached the radius $u=2.5\, \rm km$ (left) and $u=5\, \rm km$ (right), indicated in black. Lower row: The non-Markovian {\rm SIG}-model with uniform waiting times on $[40\, \rm sec,120\,\rm sec]$ for both infected and immune devices, stopped at the time the malware has reached the radius $u=2.5\, \rm km$ (left) and $u=5\, \rm km$ (right), indicated in black.}
\label{Pix-WK-Stop-Exp}
\end{figure}

Like in the previous sections, also the white-knight process can be formulated where exponential waiting times are replaced by more general renewal processes, which we call the {\em non-Markovian {\rm SIG}-model}, see Figures~\ref{Pix-WK-Stop-Exp} for illustrations. 

After having set up our models, in the next section we initiate the investigation of the models with respect to a decisive quantity, the speed of propagation. 

\subsection{The speed of malware propagation}\label{SpeedOfMalware}
One of the most important characteristics of any malware propagation model is the speed with which the malware spreads in space. More precisely, we want to assume that the malware is present only at the origin at initial time and consider balls $B_u(o)$ of radius $u\ge 0$ centered at the origin $o$. Then, we define the {\em speed} of infection spreading as
$$\limsup_{u\uparrow\infty}u\E^o[1/\tau_u]=\a,$$
where $\tau_u=\inf\{t>0:\, I(t)\not\subset B_u(o)\}$ is the hitting time of the infection at the distance $u$. Here, $\P^o$ is the Palm distribution of the Cox point process based on the street system $S$ given by a Poisson--Voronoi tessellation with edge-length intensity $\gamma$, see for example~\cite{baccelli2009stochastic1} for background on Palm distributions. 

\section{Results}\label{results}
\subsection{Simulations for the {\rm SI} models}
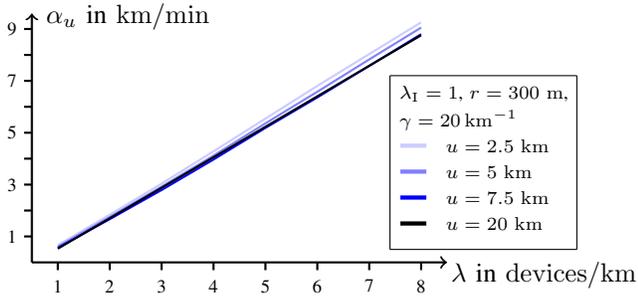
\begin{figure}[!htpb]
\centering
\begin{tikzpicture} [xscale=0.69,yscale=0.345]
		\draw[thick, ->] (0.5,0) -- (8.5,0);
		\draw[thick, ->] (0.5,0) -- (0.5,9.6);
		\coordinate[label = 0 : $\a_u$ in $\rm km/min$](a) at  (0.6,9.5);
		\coordinate[label = 0 :$\la$ in $\rm devices/km$ ](a) at  (8.4,-0.5);
\foreach \x in {1,2,3,4,5,6,7,8}
 \node[anchor=north] at (\x,-0.3) {\scriptsize{\x}} ;

\foreach \x in {1,2,3,4,5,6,7,8}
\draw[thick] (\x,0) -- (\x,-0.3);

\foreach \y in {1,3,5,7,9}
 \node[anchor=west] at (-0.15,\y) {\scriptsize{\y}};

\foreach \y in {1,2,3,4,5,6,7,8,9}
 \draw[thick] (0.35,\y) -- (0.5,\y);

\coordinate (a) at (7.4,6.5);

\coordinate (w) at (3.6,0);
\coordinate(linespaceing) at (0.0,1.0);
\coordinate(legendlength) at (0.5,0);
\coordinate(epsilonx) at (0.2,0);
\coordinate (h) at ($6.0*(linespaceing)$);

\draw[thin] ($ (a)+0.7*(linespaceing)  $) -- ($ (a) - (h) $ );
\draw[thin] ($ (a) - (h)  $) -- ($ (a) - (h) + (w) $ );
\draw[thin] ($ (a) - (h) + (w) $ ) -- ($ (a) + (w) +0.7*(linespaceing)$  );
\draw[thin] ($ (a) + (w) +0.7*(linespaceing)$) -- ($ (a)+0.7*(linespaceing)$ );

\node[anchor=west] at ($(a)$) {\scriptsize $\laI=1$, $r=300 \ \rm{ m}$,} ;
\node[anchor=west] at ($(a)-(linespaceing)$) {\scriptsize $\gamma=20\, \rm km^{-1}$} ;

\draw[ultra thick, blue!20]($(a)+(epsilonx)-2*(linespaceing)$) -- ($(a)+(epsilonx)-2*(linespaceing)+(legendlength)$);
\node[anchor=west] at ($(a)-2*(linespaceing)+(legendlength)+2*(epsilonx)$) {\scriptsize$u=2.5 \  \rm{km}$} ;

\draw[ultra thick, blue!50]($(a)+(epsilonx)-3*(linespaceing)$) -- ($(a)+(epsilonx)-3*(linespaceing)+(legendlength)$);
\node[anchor=west] at ($(a)-3*(linespaceing)+(legendlength)+2*(epsilonx)$) {\scriptsize$u=5 \  \rm{km}$};

\draw[ultra thick, blue]($(a)+(epsilonx)-4*(linespaceing)$) -- ($(a)+(epsilonx)-4*(linespaceing)+(legendlength)$);
\node[anchor=west] at ($(a)-4*(linespaceing)+(legendlength)+2*(epsilonx)$) {\scriptsize$u=7.5 \  \rm{km}$};

\draw[ultra thick, black]($(a)+(epsilonx)-5*(linespaceing)$) -- ($(a)+(epsilonx)-5*(linespaceing)+(legendlength)$);
\node[anchor=west] at ($(a)-5*(linespaceing)+(legendlength)+2*(epsilonx)$) {\scriptsize$u=20 \  \rm{km}$};

\draw [thick, blue!20, xshift=0cm]  
		plot [tension=0.8] coordinates 
 {(1.0,0.6642066903929107)(2.0,1.8399207402019946)(3.0,3.0390584169308847)(4.0,4.2834578023721175)(5.0,5.54375835046085)(6.0,6.806073668165853)(7.0,8.03349261969386)(8.0,9.25775917862971)};
\draw [thick, blue!50, xshift=0cm]  
		plot [tension=0.8] coordinates 
{(1.0,0.6062004681531986)(2.0,1.7521190883579902)(3.0,2.9211648900566187)(4.0,4.117115678852305)(5.0,5.360490192584205)(6.0,6.580008992767778)(7.0,7.826714662657043)(8.0,9.054184902257148)};	
\draw [thick, blue, xshift=0cm]  
		plot [tension=0.8] coordinates 
{(1.0,0.55944375528374)(2.0,1.6678400547602374)(3.0,2.7913398628073876)(4.0,3.9676719832998177)(5.0,5.1861882084364455)(6.0,6.363073970897325)(7.0,7.578679935933108)(8.0,8.799663995236587)};

\draw [thick, black, xshift=0cm]  
		plot [tension=0.8] coordinates 
{(1, 0.52740178)(8, 8.7564436)};
\end{tikzpicture}
\caption{Simulations for the speed of malware propagation for the Markovian {\rm SI}-model for growing observation windows as a function of $\la$.}
\label{Pix-Rich-Speed-GrowingVolume}
\end{figure}

%
%
%
%
%
%
We will approximate $\a$ by $\a_u=u\E^o[1/\tau_u]$ for some fixed radius $u$. In order to determine how fast the convergence is as $u$ tends to infinity, we estimate $\a_u$ for different values of $u$. As mentioned earlier, it is reasonable to expect that $u\mapsto \a_u$ is decreasing, since boundaries of larger balls are more easily disconnected from the initially infected node in the network model. Another aspect that works in the same direction is that the spread of the malware should behave almost independently in different directions. But, for the speed of the malware, only the maximal distance in one of the directions is taken into account. While for larger radii the law of large numbers averages the growth over all directions, for small radii, the different speeds have a higher variance. Therefore, the expected maximum is larger for small radii, although the expected growth speed in a given direction is independent of the direction. In Figure~\ref{Pix-Rich-Speed-GrowingVolume} we present estimated curves $\la\mapsto \a_u(\la)$ for different values of $u$, which indeed show the expected behavior. In particular, we can observe indications for convergence since the graphs start to become closer for linearly-growing radii and linear dependence of $\a_u$ on $\la$.

\begin{figure}[!htpb]
\centering
  \begin{tikzpicture} [xscale=0.63,yscale=6.3]
		\draw[thick, ->] (0.5,0) -- (9.5,0);
		\draw[thick, ->] (0.5,0) -- (0.5,0.4);
		\coordinate[label = 0 : $\a_u$ in $\rm km/min$](a) at  (0.6,0.4);
		\coordinate[label = 0 : $\la$ in $\rm devices/km$](a) at  (9.4,-0.035);
\foreach \x in {1,2,3,4,5,6,7,8,9}
 \node[anchor=north] at (\x,-0.01) {\scriptsize{\x}} ;
\foreach \x in {1,2,3,4,5,6,7,8,9}
\draw[thick] (\x,0) -- (\x,-0.015);
\foreach \y in {0.1,0.2,0.3}
 \node[anchor=east] at (0.5,\y) {\scriptsize{\y}};
\foreach \y in {0.1,0.2,0.3}
 \draw[thick] (0.35,\y) -- (0.5,\y);
%
\coordinate (a) at (7.5,0.37);

\coordinate (w) at (5.5,0);
\coordinate(linespaceing) at (0.0,0.057);
\coordinate(legendlength) at (0.5,0);
\coordinate(epsilonx) at (0.2,0);
\coordinate (h) at ($6.0*(linespaceing)$);

\draw[thin] ($ (a)+1*(linespaceing)  $) -- ($ (a) - (h) $ );
\draw[thin] ($ (a) - (h)  $) -- ($ (a) - (h) + (w) $ );
\draw[thin] ($ (a) - (h) + (w) $ ) -- ($ (a) + (w) +1*(linespaceing)$  );
\draw[thin] ($ (a) + (w) +1*(linespaceing)$) -- ($ (a)+1*(linespaceing)$ );

\node[anchor=west] at ($(a)$) {\scriptsize$[c,d]=[40,120]\,\rm sec$,} ;
\node[anchor=west] at ($(a)-(linespaceing)$) {\scriptsize$r=300 \ \rm{ m}$, $\gamma=20\, \rm km^{-1}$} ;

\draw[ultra thick, blue!20]($(a)+(epsilonx)-2*(linespaceing)$) -- ($(a)+(epsilonx)-2*(linespaceing)+(legendlength)$);
\node[anchor=west] at ($(a)-2*(linespaceing)+(legendlength)+2*(epsilonx)$) {\scriptsize$u=2.5 \  \rm{km}$} ;

\draw[ultra thick, blue!50]($(a)+(epsilonx)-3*(linespaceing)$) -- ($(a)+(epsilonx)-3*(linespaceing)+(legendlength)$);
\node[anchor=west] at ($(a)-3*(linespaceing)+(legendlength)+2*(epsilonx)$) {\scriptsize$u=5 \  \rm{km}$};

\draw[ultra thick, blue]($(a)+(epsilonx)-4*(linespaceing)$) -- ($(a)+(epsilonx)-4*(linespaceing)+(legendlength)$);
\node[anchor=west] at ($(a)-4*(linespaceing)+(legendlength)+2*(epsilonx)$) {\scriptsize$u=7.5 \  \rm{km}$};

\draw[ultra thick, black]($(a)+(epsilonx)-5*(linespaceing)$) -- ($(a)+(epsilonx)-5*(linespaceing)+(legendlength)$);
\node[anchor=west] at ($(a)-5*(linespaceing)+(legendlength)+2*(epsilonx)$) {\scriptsize$u=20 \  \rm{km}$};

\draw [thick, green, xshift=0cm]  
		plot [tension=0.8] coordinates 
{(1.0,0.18871473395401955)(2.0,0.24488844651881447)(3.0,0.27475735032265686)(4.0,0.2911275539473881)(5.0,0.3035778570134197)(6.0,0.311663808955839)(7.0,0.31953114294290513)};
\draw [thick, blue, xshift=0cm]  
		plot [tension=0.8] coordinates 
{(1.0,0.18320376526297355)(2.0,0.24382138631523723)(3.0,0.2726102192529955)(4.0,0.29019872498413146)(5.0,0.30279735827419246)(6.0,0.3113610478676684)(7.0,0.31934483950267334)};	
\draw [thick, black, xshift=0cm]  
		plot [tension=0.8] coordinates 
{(1.0,0.17522726937642574)(2.0,0.24024881143960503)(3.0,0.2700278394411707)(4.0,0.2879991683931767)(5.0,0.30057288541625027)(6.0,0.3097805568967026)(7.0,0.31767511499904827)};
\end{tikzpicture}
\caption{Simulations for the speed of malware propagation for the non-Markovian {\rm SI}-model with uniform waiting times, for growing observation windows as a function of $\la$.}
\label{Pix-Rich-Speed-GrowingVolume-NonMark}
\end{figure}
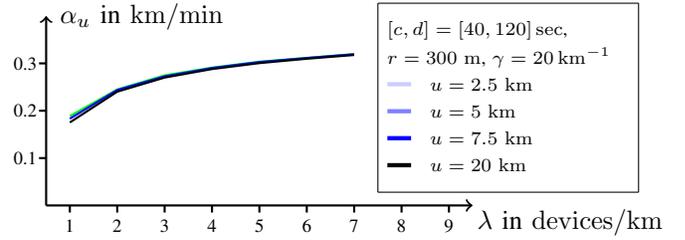

\medskip
Next we present the associated estimates for the non-Markovian dynamics, where the exponential waiting times with rate $\laI$ are replaced by uniform waiting times in the interval $[c,d]$ with $c=40\,\rm sec$ and $d=120\, \rm sec$. The results are presented in Figure~\ref{Pix-Rich-Speed-GrowingVolume-NonMark}. We can clearly observe a different behavior of the curves compared to the Markovian case. First, the convergence with respect to growing balls is faster, so that estimating $\a$ via $\a_u$ for large $u$ seems to be less important. Second, and more interesting, there is no linear dependence on $\la$ any more but rather a saturation effect for large $\la$. This saturation effect can be understood via the following observation. For denser and denser graphs, the malware is more and more able to achieve its maximal updating speed $c$. This $c$, properly scaled by $r$, then serves as an upper bound for the achievable speed.  More precisely, the infection cannot spread faster than one hop every $40\,\rm sec$. With a connection radius of $300\,\rm m$, an upper bound for the maximum speed is given by $450\, \rm m/min$, when we ignore the additional distance given by the street system. Let us note that this additional distance can be related to the {\em stretch factor} of the Cox point process, see~\cite{yao} or \cite{CaGaHiJaEnPa18}. There are two additional effects that lead to the fact that this maximum speed is not achieved for finite $\la$. First, the maximum speed can only be obtained along a straight line. On the graph however, the stretch factor encodes the shortest distance in the graph towards a large radius, asymptotically. As the number of devices increases, the difference between euclidean and graph distance encoded in the stretch factor becomes smaller. The second effect however is probably the dominating one for the shape of the curves in Figure~\ref{Pix-Rich-Speed-GrowingVolume-NonMark}. The maximal speed is achieved if at every step the minimal infection time is used. As the number of devices increase, so does the number of possible edges between devices (even quadratically). More available edges lead to a higher likeliness to sample an edge with a fast infection time.

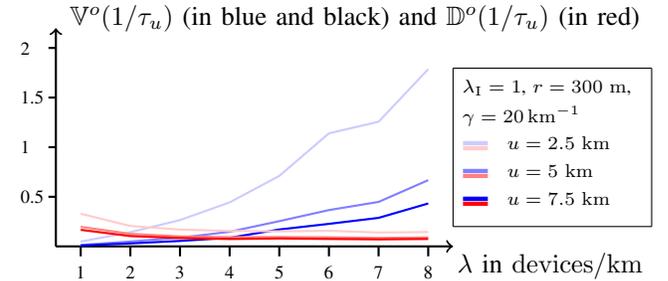
\begin{figure}[!htpb]
\centering
\begin{tikzpicture} [xscale=0.66,yscale=1.32]
		\draw[thick, ->] (0.5,0) -- (8.5,0);
		\draw[thick, ->] (0.5,0) -- (0.5,2.2);
		\coordinate[label = 0 : $\mathbb V^o(1/\tau_u)$ (in blue and black) and $\mathbb D^o(1/\tau_u)$ (in red) ](a) at  (0.6,2.3);
		\coordinate[label = 0 : $\la$ in $\rm devices/km$ ](a) at  (8.4,-0.2);

\foreach \x in {1,2,3,4,5,6,7,8}
 \node[anchor=north] at (\x,-0.1) {\scriptsize{\x}} ;

\foreach \x in {1,2,3,4,5,6,7,8}
\draw[thick] (\x,0) -- (\x,-0.1);

\foreach \y in {0.5,1,1.5,2}
 \node[anchor=west] at (-0.4,\y) {\scriptsize{\y}};

\foreach \y in {0.5,1,1.5,2}
 \draw[thick] (0.35,\y) -- (0.5,\y);

\coordinate (a) at (8.5,1.6);

\coordinate (w) at (4,0);
\coordinate(linespaceing) at (0.0,0.28);
\coordinate(legendlength) at (0.5,0);
\coordinate(epsilonx) at (0.2,0);
\coordinate (h) at ($5.0*(linespaceing)$);
\coordinate(epsilony) at (0,0.025);
\draw[thin] ($ (a)+0.7*(linespaceing)  $) -- ($ (a) - (h) $ );
\draw[thin] ($ (a) - (h)  $) -- ($ (a) - (h) + (w) $ );
\draw[thin] ($ (a) - (h) + (w) $ ) -- ($ (a) + (w) +0.7*(linespaceing)$  );
\draw[thin] ($ (a) + (w) +0.7*(linespaceing)$) -- ($ (a)+0.7*(linespaceing)$ );

\node[anchor=west] at ($(a)$) {\scriptsize$\laI=1$, $r=300 \ \rm{ m}$,} ;
\node[anchor=west] at ($(a)-(linespaceing)$) {\scriptsize$\gamma=20\, \rm km^{-1}$} ;

\draw[ultra thick, blue!20]($(a)+(epsilonx)-2*(linespaceing)$) -- ($(a)+(epsilonx)-2*(linespaceing)+(legendlength)$);
\node[anchor=west] at ($(a)-2*(linespaceing)+(legendlength)+2*(epsilonx)$) {\scriptsize$u=2.5 \  \rm{km}$} ;
\draw[ultra thick, red!20]($(a)+(epsilonx)-2*(linespaceing)-2*(epsilony)$) -- ($(a)+(epsilonx)-2*(linespaceing)+(legendlength)-2*(epsilony)$);

\draw[ultra thick, blue!50]($(a)+(epsilonx)-3*(linespaceing)$) -- ($(a)+(epsilonx)-3*(linespaceing)+(legendlength)$);
\node[anchor=west] at ($(a)-3*(linespaceing)+(legendlength)+2*(epsilonx)$) {\scriptsize$u=5 \  \rm{km}$};
\draw[ultra thick, red!50]($(a)+(epsilonx)-3*(linespaceing)-2*(epsilony)$) -- ($(a)+(epsilonx)-3*(linespaceing)+(legendlength)-2*(epsilony)$);

\draw[ultra thick, blue]($(a)+(epsilonx)-4*(linespaceing)$) -- ($(a)+(epsilonx)-4*(linespaceing)+(legendlength)$);
\node[anchor=west] at ($(a)-4*(linespaceing)+(legendlength)+2*(epsilonx)$) {\scriptsize$u=7.5 \  \rm{km}$};
\draw[ultra thick, red]($(a)+(epsilonx)-4*(linespaceing)-2*(epsilony)$) -- ($(a)+(epsilonx)-4*(linespaceing)+(legendlength)-2*(epsilony)$);

%

\draw [thick, blue!20, xshift=0cm]  
		plot [tension=0.8] coordinates 
 {(1,         0.04775562) (2,         0.1428536) (3,         0.26508681 ) (4,         0.44331122 ) (5,         0.71026567) (6,         1.13959953 ) (7,         1.25601232) (8,         1.78623555 )}; 
\draw [thick, blue!50, xshift=0cm]  
		plot [tension=0.8] coordinates 
{(1,          0.01425653) (2,           0.0490409  ) (3,         0.08695697) (4,         0.14712954) (5,          0.25420108)(6,          0.36577381)(7,          0.44815087)(8,         0.66757777)};
\draw [thick, blue, xshift=0cm]  
		plot [tension=0.8] coordinates 
{(1,         0.00870531) (2,         0.03010785) (3,         0.05482594) (4,         0.08686365) (5,         0.17016068) (6,         0.22696294) (7,         0.28731342) (8,         0.43251636)}; 


 \draw [thick, red!20, xshift=0cm]  
		plot [tension=0.8] coordinates 
 {(1,        0.3290099243) (2,        0.2054217761) (3,        0.1694162291 ) (4,         0.1554389554 ) (5,        0.1520218876) (6,         0.1568481805 ) (7,         0.1395058921) (8,         0.1443655218)}; 

 \draw [thick, red!50, xshift=0cm]  
		plot [tension=0.8] coordinates 
 {(1,       0.1969657198) (2,        0.1263908383) (3,        0.10094763 ) (4,        0.0931658754 ) (5,       0.0940554986) (6,        0.0919136084 ) (7,        0.0855327865) (8,        0.0902405168)};

\draw [thick, red, xshift=0cm]  
		plot [tension=0.8] coordinates 
 {(1,      0.1667768191) (2,        0.1040364378) (3,       0.0838842307) (4,      0.0742819467) (5,      0.0795392209) (6,      0.0748704602) (7,     0.0707268619) (8,       0.0747369172)}; 

\end{tikzpicture}

\caption{Simulations for the variances (in blue and black) and the relative deviations (in red) in the simulations of the speed of malware propagation for the Markovian {\rm SI}-model for growing observation windows. }
\label{Pix-Rich-Speed-GrowingVolume-Var}
\end{figure}

\medskip
In order to further evaluate the convergence properties of our estimates, we track the associated approximate variance 
$$\V^o[1/\tau_u]\approx n^{-2}\sum_{i=1}^n (1/\tau^{i}_u)^2-\big(n^{-1}\sum_{i=1}^n 1/\tau^{i}_u\big)^2.$$
The corresponding curves for the Markovian setting are plotted in Figure~\ref{Pix-Rich-Speed-GrowingVolume-Var} in blue. We observe a substantial decrease in variance for larger windows. This is not surprising since larger windows provide better averaging with respect to the waiting times. In other words, in smaller observation windows, less updating is involved which consequently creates more variance. At the same time, in larger observation windows it is even more unlikely to see clusters which contain the initial node where the malware starts and which are also large but do not reach the boundary of the window. Essentially, in large windows only the infinite cluster reaches the boundary of the window. 

However, for larger values of $\la$ the absolute variance increases, which is simply due to the fact that the variance measures the $L^2$ distance to larger expected values for larger $\la$. In order to see that larger values of $\la$ lead to less deviation from the mean, in Figure~\ref{Pix-Rich-Speed-GrowingVolume-Var} we also plot the relative deviation from the mean in red. The relative deviation is formally defined by 
$$\mathbb D^o[1/\tau_u]\approx \frac{\sqrt{n^{-2}\sum_{i=1}^n (1/\tau^{i}_u)^2-\big(n^{-1}\sum_{i=1}^n 1/\tau^{i}_u\big)^2}}{n^{-1}\sum_{i=1}^n 1/\tau^{i}_u}.$$
This decrease in the relative deviation is due to the fact that more nodes, encoded in larger values of $\la$, provide more updates per unit volume, which results in more averaging, i.e., less deviation. Plus, larger $\la$ refers to being deeper in the supercritical percolation regime for the network model, which also boosts the effect that we just described, namely that in large windows, for the boundary to be connected to the initial node by clusters other than the (unique) infinite one is highly unlikely.

Let us finally note that for the non-Markovian model the variances in all distances are already very small for $\la$ not too close to the critical one, therefore we omit the associated plots.

\subsection{Simulations for the {\rm SIG} models}
In this section we provide simulation results for the {\rm SIG}-models, where the counter measure is given by white knights that are able to install patches exclusively on neighboring devices that carry the malware. 
Our main objective is to present simulations for the phase diagram of survival and extinction of the malware. Here, we determine the critical intensity of white knights needed in order to eliminate the malware in the system for varying rates of the infection transmission. Let us again fix the parameters $\gamma =20\,\rm km^{-1}$, $\lambda=2\, \rm devices/km$ and $r=300\, \rm m$. 

\subsubsection{The Markovian models}
Let us start by giving an account on the Markovian models. Recall that here we can fix $\laG=1$, the rate at which patches are installed on previously infected devices, due to time rescaling. In Figure~\ref{Pix-WK-Stop-Exp} we illustrate the system stopped at two random times at which the malware has first reached the boundary of a certain prescribed centered ball.

Note that we can observe that in general the spreading of the malware can be divided into two main phases. In the first phase, as the white knights are random and maybe rare, the process has not discovered any white knights and spreads unobstructed like the {\rm SI}-models. 
In the second phase, the first white knight has been discovered and a chase-escape dynamic unfolds in which the patch starts to spread on the previously infected nodes. 

In the infinite volume, it is reasonable to assume, although not rigorously proven in our setting, that, for any value of intensity of the white knights $\rho$, there exists a {\em unique} critical infection rate $\laI^{\rm c}$ such that for $\laI>\laI^{\rm c}$ the infection survives for all times and infects infinitely-many devices, and for $\laI<\laI^{\rm c}$, the infection goes extinct on any infinite network component. Let us note that this result is available for $d$-regular trees, see~\cite{Ko05}.  In~\cite{HiJaCaWa19}, for the case where $\Lambda$ is the Lebesgue measure, we could rigorously determine areas in the phase diagram in which extinction, respectively survival, is present, but had to leave open the precise shape of the critical phase-separation line, and in fact also an area around that suspected line, due to lack of monotonicity. To determine this line via simulations in a much more realistic setting is one of the main results of this paper. 

\begin{figure}[!htpb]
\centering
 \begin{tikzpicture} [xscale=2.4,yscale=2.4]
		\draw[thick, ->] (0.5,0) -- (2.2,0);
		\draw[thick, ->] (0.5,0) -- (0.5,1);
		\coordinate[label = 0 : $\P^o\big(\tau_u<\infty|o\leftrightsquigarrow B^{\rm c}_u(o)\big)$](a) at  (0.5,1.1);
		\coordinate[label = 0 : $\laI$ ](a) at  (2.25,-0.05);
\foreach \x in  {0.8,1.0,1.2,1.4,1.6,1.8,2.0}
 \node[anchor=north] at (\x,-0.05) {\scriptsize{\x}} ;

\foreach \x in {0.8,1.0,1.2,1.4,1.6,1.8,2.0}
\draw[thick] (\x,0) -- (\x,-0.05);

\foreach \y in {0.2,0.4,0.6,0.8,1}
 \node[anchor=east] at (0.48,\y) {\scriptsize{\y}};

\foreach \y in {0.2,0.4,0.6,0.8}
 \draw[thick] (0.45,\y) -- (0.5,\y);

\coordinate (a) at (2.15,1.15);

\coordinate (w) at (1.7,0);
\coordinate(linespaceing) at (0.0,0.14);
\coordinate(legendlength) at (0.25,0);
\coordinate(epsilonx) at (0.05,0);
\coordinate (h) at ($7.8*(linespaceing)$);

 \draw[thin] ($ (a)+0.5*(linespaceing)  $) -- ($ (a) - (h) $ );
\draw[thin] ($ (a) - (h)  $) -- ($ (a) - (h) + (w) $ );
\draw[thin] ($ (a) - (h) + (w) $ ) -- ($ (a) + (w) +0.5*(linespaceing)$  );
\draw[thin] ($ (a) + (w) +0.5*(linespaceing)$) -- ($ (a)+0.5*(linespaceing) $ );

\node[anchor=west] at ($(a)$) {\scriptsize$\rho=0.5\ \rm devices/km$, $r=300 \ \rm{ m}$,} ;
\node[anchor=west] at ($(a)-(linespaceing)$) {\scriptsize$\la=2\  \rm devices/km$, $\gamma=20\ \rm{ km}^{-1}$} ;

\draw[ultra thick, blue!20]($(a)+(epsilonx)-2*(linespaceing)$) -- ($(a)+(epsilonx)-2*(linespaceing)+(legendlength)$);
\node[anchor=west] at ($(a)-2*(linespaceing)+(legendlength)+2*(epsilonx)$) {\scriptsize$u=2.5 \ \rm{ km}$} ;

\draw[ultra thick, blue!40]($(a)+(epsilonx)-3*(linespaceing)$) -- ($(a)+(epsilonx)-3*(linespaceing)+(legendlength)$);
\node[anchor=west] at ($(a)-3*(linespaceing)+(legendlength)+2*(epsilonx)$) {\scriptsize$u=5 \ \rm{ km}$};

\draw[ultra thick, blue!60]($(a)+(epsilonx)-4*(linespaceing)$) -- ($(a)+(epsilonx)-4*(linespaceing)+(legendlength)$);
\node[anchor=west] at ($(a)-4*(linespaceing)+(legendlength)+2*(epsilonx)$) {\scriptsize$u=7.5 \ \rm{ km}$};

\draw[ultra thick, blue!80]($(a)+(epsilonx)-5*(linespaceing)$) -- ($(a)+(epsilonx)-5*(linespaceing)+(legendlength)$);
\node[anchor=west] at ($(a)-5*(linespaceing)+(legendlength)+2*(epsilonx)$) {\scriptsize$u=10 \ \rm{ km}$};

\draw[ultra thick, blue]($(a)+(epsilonx)-6*(linespaceing)$) -- ($(a)+(epsilonx)-6*(linespaceing)+(legendlength)$);
\node[anchor=west] at ($(a)-6*(linespaceing)+(legendlength)+2*(epsilonx)$) {\scriptsize$u=12.5 \ \rm{ km}$}; ;

\draw[ultra thick,black]($(a)+(epsilonx)-7*(linespaceing)$) -- ($(a)+(epsilonx)-7*(linespaceing)+(legendlength)$);
\node[anchor=west] at ($(a)-7*(linespaceing)+(legendlength)+2*(epsilonx)$) {\scriptsize$u=15 \ \rm{ km}$};

\draw [thick, blue!20, xshift=0cm]  
		plot [tension=0.8] coordinates 
 {(0.8,0.07)(1.0,0.21)(1.2,0.43)(1.4,0.58)(1.6,0.66)(1.7,0.7)(1.8,0.725)(1.9,0.85)(2.0,0.855)};
\draw [thick, blue!40, xshift=0cm]  
		plot [tension=0.8] coordinates 
{(0.8,0.005)(1.0,0.025)(1.2,0.17)(1.4,0.44)(1.6,0.6)(1.7,0.685)(1.8,0.705)(1.9,0.845)(2.0,0.855)};	
\draw [thick, blue!60, xshift=0cm]  
		plot [tension=0.8] coordinates 
{(0.8,0.0)(1.0,0.0)(1.2,0.06)(1.4,0.29)(1.6,0.565)(1.7,0.675)(1.8,0.705)(1.9,0.845)(2.0,0.855)};
\draw [thick, blue!80, xshift=0cm]  
		plot [tension=0.8] coordinates 
{(0.8,0.0)(1.0,0.0)(1.2,0.025)(1.4,0.2)(1.6,0.545)(1.7,0.665)(1.8,0.705)(1.9,0.845)(2.0,0.855)};
\draw [thick, blue, xshift=0cm]  
		plot [tension=0.8] coordinates 
{(0.8,0.0)(1.0,0.0)(1.2,0.005)(1.4,0.095)(1.6,0.535)(1.7,0.66)(1.8,0.705)(1.9,0.84)(2.0,0.855)};
\draw [thick, black, xshift=0cm]  
		plot [tension=0.8] coordinates 
{(0.8,0.0)(1.0,0.0)(1.2,0.0)(1.4,0.055)(1.6,0.44)(1.7,0.63)(1.8,0.705)(1.9,0.84)(2.0,0.855)};
\end{tikzpicture}
\caption{Estimates for the survival probability, conditioned to the event that a cluster exists connecting the initially infected node at the origin to the boundary of the observation window, for varying radii and in the Markovian setting.}
\label{Pix-WK-Rho-Lambda-1d}
\end{figure}
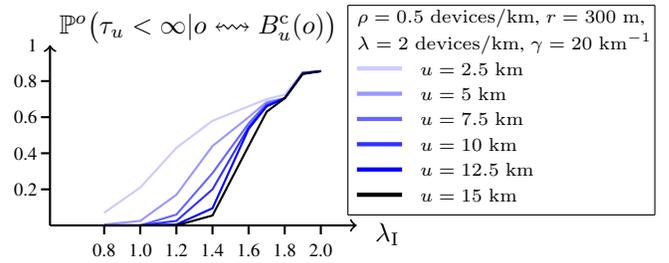

In order to determine the critical infection rate $\laI^{\rm c}$ for one given $\rho$, we have to cope with the fact that we simulate in bounded observation windows. More precisely, in the infinite domain we could simply define $\laI^{\rm c}=\inf\{\laI\colon \tilde\a>0\}$, where $\tilde\a$ is defined as $\a$ but now under the measure $\P^o$ conditioned on connectedness of the origin to infinity. However, in finite windows survival is checked only within the window and thus survival inside the window could in principle still mean extinction in any larger domain. In Figure~\ref{Pix-WK-Rho-Lambda-1d} we present estimates for the probability that the infection reaches varying distances $u$ in finite time.

As could be anticipated, the curves become more pronounced in larger observation windows, since there we see the general principle that it is highly unlikely to survive a long time and then die out. In order to have a good balance between computational complexity and accuracy of the estimates, to find the critical intensity $\laI^{\rm c}$, we simulate large windows and check for which $\laI$ the proportion of realizations in which the malware reaches the boundary of the window exceeds $60\%$. 
For this note that it is computationally cheaper to run simulations in the extinction regime than in the survival regime. This is simply because in the extinction regime the infection dies out faster and hence our stopping criterion applies earlier. Therefore, in order to obtain the critical values, we run $1000$ simulations ($10$ realizations of the propagation model on $100$ realizations of the D2D-network model) until the probability to survive (in this case for the infection to reach a distance more than $17.5\,\rm km$ from the initially infected node) is larger than $60\%$. Note that the accuracy of the estimate of the critical value depends on the chosen discretization in the infection speed $\laI$ and the number of different white-knight intensities $\rho$ that are simulated.

%

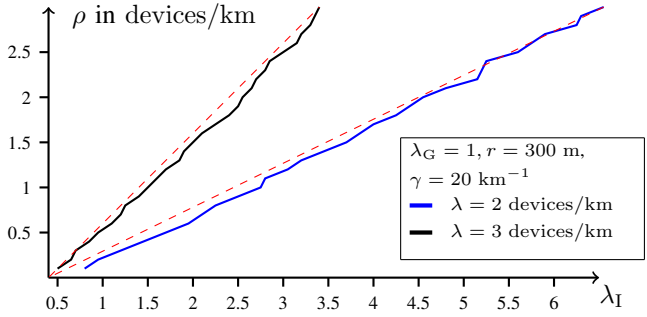
\begin{figure}[!htpb]
\centering
 \begin{tikzpicture} [xscale=1.2,yscale=1.2]
		\draw[thick, ->] (0.4,0) -- (6.5,0);
		\draw[thick, ->] (0.4,0) -- (0.4,3);
		\coordinate[label = 180 : $\rho$ in $\rm devices/km$](a) at  (2.8,2.9);
		\coordinate[label = 0 : $\laI$ ](a) at  (6.4,-0.2);
\foreach \x in  {0.5,1,1.5,2,2.5,3,3.5,4,4.5,5,5.5,6}
 \node[anchor=north] at (\x,-0.1) {\scriptsize{\x}} ;

\foreach \x in {0.5,1,1.5,2,2.5,3,3.5,4,4.5,5,5.5,6}
\draw[thick] (\x,0) -- (\x,-0.1);

\foreach \y in {0.5,1,1.5,2,2.5}
 \node[anchor=east] at (0.31,\y) {\scriptsize{\y}};

\foreach \y in {0.5,1,1.5,2,2.5}
 \draw[thick] (0.30,\y) -- (0.4,\y);

\coordinate (a) at (4.3,1.4);

\coordinate (w) at (2.7,0);
\coordinate(linespaceing) at (0.0,0.3);
\coordinate(legendlength) at (0.25,0);
\coordinate(epsilonx) at (0.1,0);
\coordinate (h) at ($4.0*(linespaceing)$);

 \draw[thin] ($ (a)+0.5*(linespaceing)  $) -- ($ (a) - (h) $ );
\draw[thin] ($ (a) - (h)  $) -- ($ (a) - (h) + (w) $ );
\draw[thin] ($ (a) - (h) + (w) $ ) -- ($ (a) + (w) +0.5*(linespaceing)$  );
\draw[thin] ($ (a) + (w) +0.5*(linespaceing)$) -- ($ (a)+0.5*(linespaceing) $ );

\node[anchor=west] at ($(a)$) {\scriptsize$\laG=1, r=300 \ \rm{ m}$,} ;
\node[anchor=west] at ($(a)-(linespaceing)$) {\scriptsize$\gamma=20\ \rm{ km}^{-1}$} ;

\draw[ultra thick, blue]($(a)+(epsilonx)-2*(linespaceing)$) -- ($(a)+(epsilonx)-2*(linespaceing)+(legendlength)$);
\node[anchor=west] at ($(a)-2*(linespaceing)+(legendlength)+2*(epsilonx)$) {\scriptsize$\la = 2 \  \rm{devices/km}$} ;

\draw[ultra thick, black]($(a)+(epsilonx)-3*(linespaceing)$) -- ($(a)+(epsilonx)-3*(linespaceing)+(legendlength)$);
\node[anchor=west] at ($(a)-3*(linespaceing)+(legendlength)+2*(epsilonx)$) {\scriptsize$\la = 3\ \rm{devices/km}$};

\draw [thick, blue, xshift=0cm]  
		plot [tension=0.8] coordinates 
{(0.7999999999999999,0.1)(0.95,0.2)(1.2000000000000002,0.3)(1.4500000000000004,0.4)(1.7000000000000006,0.5)(1.9500000000000008,0.6)(2.100000000000001,0.7)(2.250000000000001,0.7999999999999999)(2.500000000000001,0.8999999999999999)(2.750000000000001,0.9999999999999999)(2.8000000000000007,1.0999999999999999)(3.0500000000000007,1.2)(3.2000000000000006,1.3)(3.4500000000000006,1.4)(3.7000000000000006,1.5)(3.8500000000000005,1.5999999999999999)(4.0,1.7)(4.249999999999999,1.8)(4.399999999999999,1.9)(4.549999999999998,2.0)(4.799999999999997,2.0999999999999996)(5.149999999999996,2.1999999999999997)(5.199999999999996,2.3)(5.249999999999996,2.4)(5.599999999999994,2.5)(5.749999999999994,2.6)(5.899999999999993,2.6999999999999997)(6.249999999999992,2.8)(6.299999999999992,2.9)(6.549999999999991,3.0)};

\draw [thick, black, xshift=0cm]  
		plot [tension=0.8] coordinates 
{(0.49999999999999994,0.1)(0.65,0.2)(0.7000000000000001,0.3)(0.8500000000000002,0.4)(0.9500000000000003,0.5)(1.1000000000000003,0.6)(1.2000000000000004,0.7)(1.2500000000000004,0.7999999999999999)(1.4000000000000006,0.8999999999999999)(1.5000000000000007,0.9999999999999999)(1.6000000000000008,1.0999999999999999)(1.7000000000000008,1.2)(1.850000000000001,1.3)(1.900000000000001,1.4)(2.000000000000001,1.5)(2.1000000000000005,1.5999999999999999)(2.25,1.7)(2.3999999999999995,1.8)(2.499999999999999,1.9)(2.549999999999999,2.0)(2.6499999999999986,2.0999999999999996)(2.6999999999999984,2.1999999999999997)(2.799999999999998,2.3)(2.849999999999998,2.4)(2.9999999999999973,2.5)(3.149999999999997,2.6)(3.1999999999999966,2.6999999999999997)(3.2999999999999963,2.8)(3.349999999999996,2.9)(3.399999999999996,3.0)};

\draw [thin, dashed, red, xshift=0cm]  
		plot [tension=0.8] coordinates 
{(6.549999999999991,3.0)(0.4,0)(3.399999999999996,3.0)};		
		
\end{tikzpicture}
\caption{Estimated phase-separating curve for the survival of the malware in the Markovian {\rm SIG}-model for two values of the linear device intensity. The dashed red lines indicate possible linear relationships. Below the line the malware is in the survival regime. Above the line, the malware becomes extinct.}
\label{WK_Rho_Lambda}
\end{figure}

Performing this step for multiple choices of $\rho$, we arrive at the simulated phase diagram presented in Figure~\ref{WK_Rho_Lambda}. 
Let us mention that we simulated the critical infection speed for white-knight intensities ranging from $0.1$ to $3$ in $0.1$ increments, i.e., $30$ values of $\rho$ and increased values of $\laI$ in steps of size $0.05$ from $0$ to $7$. This would in principle lead to around $30\times 140$ data points. However, we can use the property that for an increasing number of white knights, the critical infection rate has to increase as it becomes harder to spread. Therefore, new simulations for increasing $\rho$ can use the previous critical value as a lower bound. This cuts down the number of simulations to roughly $30+140$.

The simulated phase-separating line resembles a straight line, indicating a linear dependence in the function $\rho\mapsto\laI(\rho)$. One way to understand this effect may be to consider an infected device surrounded by white knights. In order for the malware to escape before the infected device receives the patch from the surrounding white knights, it has to be faster than the minimal patching event coming from the white knights. But for twice as many white knights, the first patch attempt is twice as fast and thus the malware should also be twice as fast in order to escape. This idea may serve as a rough guideline for the understanding of the linear relationship.

\subsubsection{The non-Markovian models}
The difference to the previous section is that now infection and immunization attempts are performed after non-exponential waiting times. As mentioned before we choose for the infection the uniform distribution $[c,d]$ with $c=40\,\rm sec$ and $d=120\,\rm sec$. Similarly, the waiting times for the patches are given by $[c,g]$ with the same $c=40\,\rm sec$ and a $g>c$, which is potentially different from $d$. This allows us to incorporate in our simulations the situation in which the patches can be installed faster or slower than the malware transmits. In Figure~\ref{Pix-WK-Stop-Exp} we illustrate the system stopped at two random times at which the malware has first reached the boundary of a certain prescribed centered ball, where we choose $g=d$.

As in the Markovian case, in order to simulate the phase diagram, we start by estimating the system for fixed white-knight intensity. Note that our propagation model has now the three parameters $c,d,g$ instead of the two parameters $\laI,\laG$. By invariance with respect to time rescaling we can eliminate one of the parameters. In order to create a setting which makes it easier to compare the non-Markovian case to the Markovian case, we decided to keep $c$ and $d$ fixed and vary $g$, since then again, for larger $g$, the probability for survival grows. 
As in Figure~\ref{Pix-WK-Rho-Lambda-1d}, also in the non-Markovian case, 
%
for larger $u$, the curves become more pronounced.

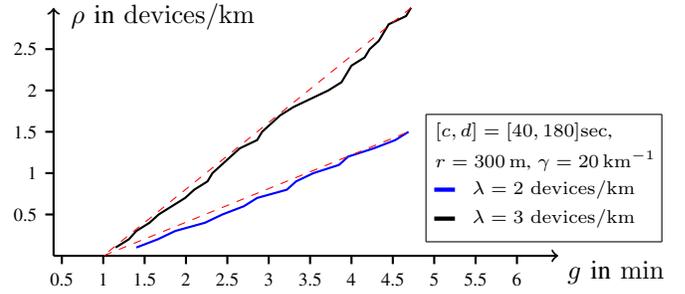
\begin{figure}[!htpb]
\centering
 \begin{tikzpicture} [xscale=1.1,yscale=1.1]
		\draw[thick, ->] (0.4,0) -- (6.5,0);
		\draw[thick, ->] (0.4,0) -- (0.4,3);
		\coordinate[label = 0 : $\rho$ in $\rm devices/km$](a) at  (0.5,2.9);
		\coordinate[label = 0 : $g$ in $\rm min$ ](a) at  (6.5,-0.2);
\foreach \x in  {0.5,1,1.5,2,2.5,3,3.5,4,4.5,5,5.5,6}
 \node[anchor=north] at (\x,-0.1) {\scriptsize{\x}} ;

\foreach \x in {0.5,1,1.5,2,2.5,3,3.5,4,4.5,5,5.5,6}
\draw[thick] (\x,0) -- (\x,-0.1);

\foreach \y in {0.5,1,1.5,2,2.5}
 \node[anchor=east] at (0.31,\y) {\scriptsize{\y}};

\foreach \y in {0.5,1,1.5,2,2.5}
 \draw[thick] (0.30,\y) -- (0.4,\y);

\coordinate (a) at (4.9,1.5);

\coordinate (w) at (2.85,0);
\coordinate(linespaceing) at (0.0,0.35);
\coordinate(legendlength) at (0.25,0);
\coordinate(epsilonx) at (0.1,0);
\coordinate (h) at ($3.8*(linespaceing)$);

\draw[thin] ($ (a)+0.6*(linespaceing)  $) -- ($ (a) - (h) $ );
\draw[thin] ($ (a) - (h)  $) -- ($ (a) - (h) + (w) $ );
\draw[thin] ($ (a) - (h) + (w) $ ) -- ($ (a) + (w) +0.6*(linespaceing)$  );
\draw[thin] ($ (a) + (w) +0.6*(linespaceing)$) -- ($ (a)+0.6*(linespaceing) $ );

\node[anchor=west] at ($(a)$) {\scriptsize$[c,d] = [40,180]\rm sec$,} ;
\node[anchor=west] at ($(a)-(linespaceing)$) {\scriptsize$r=300\, \rm m$,  $\gamma=20\,\rm km^{-1}$} ;

\draw[ultra thick, blue]($(a)+(epsilonx)-2*(linespaceing)$) -- ($(a)+(epsilonx)-2*(linespaceing)+(legendlength)$);
\node[anchor=west] at ($(a)-2*(linespaceing)+(legendlength)+2*(epsilonx)$) {\scriptsize$\la = 2 \  \rm{devices/km}$} ;

\draw[ultra thick, black]($(a)+(epsilonx)-3*(linespaceing)$) -- ($(a)+(epsilonx)-3*(linespaceing)+(legendlength)$);
\node[anchor=west] at ($(a)-3*(linespaceing)+(legendlength)+2*(epsilonx)$) {\scriptsize$\la = 3\ \rm{devices/km}$};

\draw [thick, black, xshift=0cm]  
		plot [tension=0.8] coordinates 
{(1.1500000000000001,0.1)(1.3100000000000003,0.2)(1.4000000000000004,0.3)(1.5600000000000005,0.4)(1.6700000000000006,0.5)(1.8300000000000007,0.6)(1.9900000000000009,0.7)(2.1000000000000005,0.8)(2.26,0.9)(2.3199999999999994,1.0)(2.4299999999999985,1.1)(2.5399999999999983,1.2)(2.6499999999999977,1.3)(2.8599999999999968,1.4)(2.9199999999999964,1.5)(3.029999999999996,1.6)(3.1399999999999952,1.7)(3.2999999999999945,1.8)(3.5099999999999936,1.9)(3.7199999999999923,2.0)(3.879999999999992,2.1)(3.9399999999999915,2.2)(3.999999999999991,2.3)(4.15999999999999,2.4)(4.21999999999999,2.5)(4.329999999999989,2.6)(4.389999999999989,2.7)(4.449999999999989,2.8)(4.659999999999988,2.9)(4.719999999999987,3.0)};

\draw [thick, blue, xshift=0cm]  
		plot [tension=0.8] coordinates 		
{(1.4000000000000004,0.1)(1.6600000000000006,0.2)(1.8700000000000008,0.3)(2.23,0.4)(2.439999999999999,0.5)(2.699999999999998,0.6)(2.859999999999997,0.7)(3.2199999999999958,0.8)(3.329999999999995,0.9)(3.5399999999999943,1.0)(3.849999999999993,1.1)(3.9599999999999924,1.2)(4.269999999999992,1.3)(4.5299999999999905,1.4)(4.68999999999999,1.5)};

\draw [thin, dashed, red, xshift=0cm]
		plot [tension=0.8] coordinates 
{(4.719999999999987,3.0)(1,0)(4.68999999999999,1.5)};

\end{tikzpicture}
\caption{Estimated phase diagram for the survival of the malware in the non-Markovian {\rm SIG}-model for two values of the linear device intensity. The dashed red lines indicate possible linear relationships. Above the line the malware is in the survival regime. Below the line, the malware becomes extinct.}
\label{Pix-WK-Rho-Lambda-NonMark}
\end{figure}

Finally using the same approach as for the Markovian model we obtain the estimated phase diagram for the non-Markovian case, which we present in Figure~\ref{Pix-WK-Rho-Lambda-NonMark}. 

Again, the simulated phase-separating line resembles a straight line, indicating a linear dependence in the function $\rho\mapsto g(\rho)$. This prediction needs to be further investigated in future research. 

\subsubsection{The dependence on the street system}
In this section we analyze the influence of the intensity of the street system on the critical infection speed $\laI^{\rm c}$ with respect to the intensity of white knights $\rho$.
In order to better compare the critical behavior of the system, we keep the spatial intensity of devices fixed, i.e., as $\gamma$ varies, we set $\la=c/ \gamma$. Recall that as $\gamma \rightarrow \infty$ and $\la\gamma=c$, the distribution of devices converges towards a Poisson point process with intensity $c$, see for example~\cite{HiJaCa19}. 

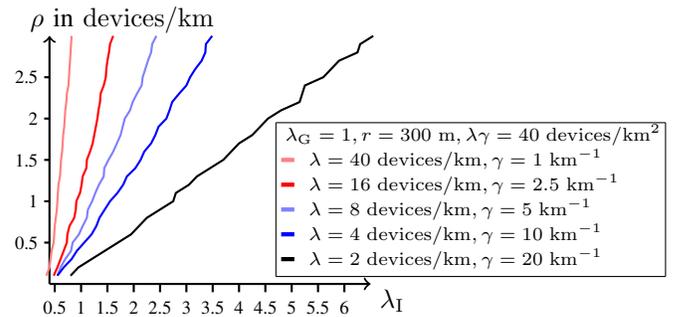
\begin{figure}[!htpb]
\centering
 \begin{tikzpicture} [xscale=0.7,yscale=1.1]
		\draw[thick, ->] (0.4,0) -- (6.5,0);
		\draw[thick, ->] (0.4,0) -- (0.4,3);
		\coordinate[label = 180 : $\rho$ in $\rm devices/km$](a) at  (3.7,3.2);
		\coordinate[label = 0 : $\laI$ ](a) at  (6.5,-0.2);
\foreach \x in  {0.5,1,1.5,2,2.5,3,3.5,4,4.5,5,5.5,6}
 \node[anchor=north] at (\x,-0.1) {\scriptsize{\x}} ;

\foreach \x in {0.5,1,1.5,2,2.5,3,3.5,4,4.5,5,5.5,6}
\draw[thick] (\x,0) -- (\x,-0.1);

\foreach \y in {0.5,1,1.5,2,2.5}
 \node[anchor=east] at (0.36,\y) {\scriptsize{\y}};

\foreach \y in {0.5,1,1.5,2,2.5}
 \draw[thick] (0.30,\y) -- (0.4,\y);

\coordinate (a) at (4.7,1.8);

\coordinate (w) at (7.4,0);
\coordinate(linespaceing) at (0.0,0.3);
\coordinate(legendlength) at (0.25,0);
\coordinate(epsilonx) at (0.1,0);
\coordinate (h) at ($5.7*(linespaceing)$);

 \draw[thin] ($ (a)+0.5*(linespaceing)  $) -- ($ (a) - (h) $ );
\draw[thin] ($ (a) - (h)  $) -- ($ (a) - (h) + (w) $ );
\draw[thin] ($ (a) - (h) + (w) $ ) -- ($ (a) + (w) +0.5*(linespaceing)$  );
\draw[thin] ($ (a) + (w) +0.5*(linespaceing)$) -- ($ (a)+0.5*(linespaceing) $ );

\node[anchor=west] at ($(a)$) {\scriptsize$\laG=1, r=300 \ \rm{ m}, \la\gamma = 40 \rm{\  devices/km^2}$} ;

\draw[ultra thick, red!50]($(a)+(epsilonx)-(linespaceing)$) -- ($(a)+(epsilonx)-(linespaceing)+(legendlength)$);
\node[anchor=west] at ($(a)-(linespaceing)+(legendlength)+2*(epsilonx)$) {\scriptsize$\la = 40 \rm{\ devices/km}  , \gamma = 1\  \rm{km}^{-1}  $} ;

\draw[ultra thick, red]($(a)+(epsilonx)-2*(linespaceing)$) -- ($(a)+(epsilonx)-2*(linespaceing)+(legendlength)$);
\node[anchor=west] at ($(a)-2*(linespaceing)+(legendlength)+2*(epsilonx)$) {\scriptsize$\la = 16 \rm{\ devices/km}  , \gamma = 2.5\  \rm{km}^{-1}  $} ;

\draw[ultra thick, blue!50]($(a)+(epsilonx)-3*(linespaceing)$) -- ($(a)+(epsilonx)-3*(linespaceing)+(legendlength)$);
\node[anchor=west] at ($(a)-3*(linespaceing)+(legendlength)+2*(epsilonx)$) {\scriptsize$\la = 8 \rm{\ devices/km}  , \gamma = 5\  \rm{km}^{-1}  $} ;

\draw[ultra thick, blue]($(a)+(epsilonx)-4*(linespaceing)$) -- ($(a)+(epsilonx)-4*(linespaceing)+(legendlength)$);
\node[anchor=west] at ($(a)-4*(linespaceing)+(legendlength)+2*(epsilonx)$) {\scriptsize$\la = 4 \rm{\ devices/km}  , \gamma = 10\  \rm{km}^{-1}  $} ;

\draw[ultra thick, black]($(a)+(epsilonx)-5*(linespaceing)$) -- ($(a)+(epsilonx)-5*(linespaceing)+(legendlength)$);
\node[anchor=west] at ($(a)-5*(linespaceing)+(legendlength)+2*(epsilonx)$) {\scriptsize$\la = 2 \rm{\ devices/km}  , \gamma = 20\  \rm{km}^{-1}  $} ;

\draw [thick, red!50, xshift=0cm]  
		plot [tension=0.8] coordinates 
{(0.34,0.1)(0.4000000000000001,0.2)(0.4200000000000001,0.3)(0.4400000000000001,0.4)(0.48000000000000015,0.5)(0.49000000000000016,0.6)(0.5000000000000001,0.7)(0.5200000000000001,0.7999999999999999)(0.5400000000000001,0.8999999999999999)(0.5500000000000002,0.9999999999999999)(0.5600000000000002,1.0999999999999999)(0.5700000000000002,1.2)(0.6000000000000002,1.3)(0.6100000000000002,1.4)(0.6200000000000002,1.5)(0.6400000000000002,1.5999999999999999)(0.6500000000000002,1.7)(0.6600000000000003,1.8)(0.6800000000000003,1.9)(0.6900000000000003,2.0)(0.7000000000000003,2.0999999999999996)(0.7200000000000003,2.1999999999999997)(0.7300000000000003,2.3)(0.7600000000000003,2.4)(0.7700000000000004,2.5)(0.7800000000000004,2.6)(0.7900000000000004,2.6999999999999997)(0.8000000000000004,2.8)(0.8100000000000004,2.9)(0.8200000000000004,3.0)};

\draw [thick, red, xshift=0cm]  
		plot [tension=0.8] coordinates 		
{(0.48000000000000015,0.1)(0.5500000000000002,0.2)(0.6100000000000002,0.3)(0.6700000000000003,0.4)(0.7300000000000003,0.5)(0.7400000000000003,0.6)(0.7800000000000004,0.7)(0.8700000000000004,0.7999999999999999)(0.9000000000000005,0.8999999999999999)(0.9100000000000005,0.9999999999999999)(0.9800000000000005,1.0999999999999999)(1.0100000000000005,1.2)(1.0300000000000005,1.3)(1.1000000000000005,1.4)(1.1100000000000005,1.5)(1.1700000000000006,1.5999999999999999)(1.2200000000000006,1.7)(1.2500000000000007,1.8)(1.2800000000000007,1.9)(1.3100000000000007,2.0)(1.3300000000000007,2.0999999999999996)(1.3600000000000008,2.1999999999999997)(1.3700000000000008,2.3)(1.4300000000000008,2.4)(1.4700000000000009,2.5)(1.4800000000000009,2.6)(1.5000000000000009,2.6999999999999997)(1.530000000000001,2.8)(1.550000000000001,2.9)(1.610000000000001,3.0)};

\draw [thick, blue!50, xshift=0cm]  
		plot [tension=0.8] coordinates 
{(0.5500000000000002,0.1)(0.6100000000000002,0.2)(0.7000000000000003,0.3)(0.8100000000000004,0.4)(0.8400000000000004,0.5)(0.9600000000000005,0.6)(1.0300000000000005,0.7)(1.1100000000000005,0.7999999999999999)(1.1300000000000006,0.8999999999999999)(1.2100000000000006,0.9999999999999999)(1.2700000000000007,1.0999999999999999)(1.3400000000000007,1.2)(1.4300000000000008,1.3)(1.4400000000000008,1.4)(1.520000000000001,1.5)(1.610000000000001,1.5999999999999999)(1.680000000000001,1.7)(1.750000000000001,1.8)(1.7900000000000011,1.9)(1.8300000000000012,2.0)(1.9300000000000013,2.0999999999999996)(1.9800000000000013,2.1999999999999997)(2.06,2.3)(2.149999999999998,2.4)(2.1699999999999977,2.5)(2.2199999999999966,2.6)(2.2699999999999956,2.6999999999999997)(2.3299999999999943,2.8)(2.349999999999994,2.9)(2.429999999999992,3.0)};

\draw [thick, blue, xshift=0cm]  
		plot [tension=0.8] coordinates 
{(0.5500000000000002,0.1)(0.6700000000000003,0.2)(0.8200000000000004,0.3)(0.9200000000000005,0.4)(1.0600000000000005,0.5)(1.2000000000000006,0.6)(1.2900000000000007,0.7)(1.3500000000000008,0.7999999999999999)(1.4500000000000008,0.8999999999999999)(1.550000000000001,0.9999999999999999)(1.690000000000001,1.0999999999999999)(1.8000000000000012,1.2)(1.8700000000000012,1.3)(2.0300000000000007,1.4)(2.1299999999999986,1.5)(2.199999999999997,1.5999999999999999)(2.3299999999999943,1.7)(2.4099999999999926,1.8)(2.4699999999999913,1.9)(2.619999999999988,2.0)(2.679999999999987,2.0999999999999996)(2.7299999999999858,2.1999999999999997)(2.859999999999983,2.3)(2.99999999999998,2.4)(3.0699999999999785,2.5)(3.1699999999999764,2.6)(3.289999999999974,2.6999999999999997)(3.3599999999999723,2.8)(3.369999999999972,2.9)(3.4899999999999696,3.0)};

\draw [thick, black, xshift=0cm]  
		plot [tension=0.8] coordinates 
{(0.7999999999999999,0.1)(0.95,0.2)(1.2000000000000002,0.3)(1.4500000000000004,0.4)(1.7000000000000006,0.5)(1.9500000000000008,0.6)(2.100000000000001,0.7)(2.250000000000001,0.7999999999999999)(2.500000000000001,0.8999999999999999)(2.750000000000001,0.9999999999999999)(2.8000000000000007,1.0999999999999999)(3.0500000000000007,1.2)(3.2000000000000006,1.3)(3.4500000000000006,1.4)(3.7000000000000006,1.5)(3.8500000000000005,1.5999999999999999)(4.0,1.7)(4.249999999999999,1.8)(4.399999999999999,1.9)(4.549999999999998,2.0)(4.799999999999997,2.0999999999999996)(5.149999999999996,2.1999999999999997)(5.199999999999996,2.3)(5.249999999999996,2.4)(5.599999999999994,2.5)(5.749999999999994,2.6)(5.899999999999993,2.6999999999999997)(6.249999999999992,2.8)(6.299999999999992,2.9)(6.549999999999991,3.0)};

\end{tikzpicture}
\caption{Estimated phase-separating curve for the survival of the malware in the Markovian {\rm SIG}-model for two values of the linear device intensity. Below the line the malware is in the survival regime. Above the line, the malware becomes extinct.}
\label{WK_Rho_Lambda-varGamma}
\end{figure}

In Figure~\ref{WK_Rho_Lambda-varGamma} we present critical curves in the $\laI$-$\rho$-plane corresponding to different values of $\gamma$ where we fix $\lambda$ such that $\la\g=40\rm{\ devices / km}^ 2$. As in Figure~\ref{Pix-WK-Rho-Lambda-NonMark}, it is reasonable to believe that the curves are indeed linear and perturbations are due to simulation errors. We can observe a clear monotonicity, where for larger values of $\gamma$, for fixed $\laI$, the critical intensity of white knights decreases. Note however that there is no reason to believe that the decrease in the gradient of the curves is linear in $\gamma$. 
We believe that the average degree of the devices in the connection graph is the crucial factor for the gradient of the critical curves. If this assertion is valid, then it can help to explain the decreasing gradient of the critical curves. Indeed, note that for sparse street systems and a fixed spatial intensity of devices, the degree of the system is large, since many devices must be positioned on few streets, with empty space around them. On the other hand, for dense streets, the degree is similar to the degree of the planar Poisson point process. 
In this context, let us mention that, if we fix the connectivity threshold $r$ and the linear intensity of devices $\la$ and decrease the street intensity, the number of devices and therefore the average degree of device decreases. For $\gamma \rightarrow 0$, the limiting rural environment, this would even lead to disconnected graphs. This is the main reason why we couple $\la$ to $\gamma$.

\section{Conclusion}
Based on a random configuration of devices placed in an urban street system of Poisson--Voronoi type, we first studied the spreading of some malware through the system without restriction. The malware is initially at the origin and infects neighboring devices after iid random waiting times. In the Markovian setting of exponential waiting times, we can clearly observe a linear dependence of the propagation speed on the device intensity. Here, the speed is measured in terms of the malware reaching large distances. We support our numerical findings by tests on the convergence in growing observation windows and estimates on the deviations. In the non-Markovian setting, where the waiting times are uniform in a compact time-interval strictly bounded away from zero, the picture changes substantially. The set of infected devices becomes much more ball-like in space and the linear dependence of the infection speed on the device intensity is replaced by a saturation effect, where more devices do not increase the speed any more. 

Finally, we augmented the system by introducing a random set of white knights with intensity $\rho$ at initial time. We simulated the critical phase-separating line of survival and extinction of the malware in the plane of $\rho$ versus the infection rate $\lambda_{\rm I}$, for different device intensities. Again, we support our findings by tests on the convergence in growing sampling windows. The main observation is that there seems to be a linear relation between $\lambda_{\rm I}$ and $\rho$, both in the Markovian and in the non-Markovian setting.  

This leads to a number of open questions, subject to future investigations. For example: How fast is the convergence of the malware propagation speed for growing sampling windows? Is there a shape theorem for the set of infected devices? Can we determine precisely the phase diagram of survival and extinction? What is the dependence of the critical curve of survival and extinction on other model parameters, for example the device intensity? 
\section*{Acknowledgements}
This research was supported by Orange S.A.~grant CRE I07263 as well as the German Research Foundation under Germany's Excellence Strategy MATH+: The Berlin Mathematics Research Center, EXC-2046/1 project ID: 390685689. We thank A.~Boubaya for his contributions to the text. 

\bibliographystyle{IEEEtran}
\bibliography{Malware}
\end{document}